\documentclass[11pt,a4paper]{scrartcl}

\usepackage{LCVision}

\usepackage{LCVision_definitions}
\usepackage{titlepic}
\usepackage[font=small,labelfont=bf,tableposition=top]{caption}

\usepackage{xpatch}
\makeatletter
\xpatchcmd\@HepConStyle
 {\edef\@upcode{\updefault}}
 {\ifdefined\shapedefault\edef\@upcode{\shapedefault}\else\edef\@upcode{\updefault}\fi}
 {}{}
\makeatother

\RedeclareSectionCommand[
  beforeskip=1.\baselineskip,
  afterskip=.5\baselineskip]{subsection}
\RedeclareSectionCommand[
  beforeskip=1.\baselineskip,
  afterskip=.5\baselineskip]{subsubsection}

\usepackage{todonotes}
\usepackage{xpatch}
\makeatletter
\xpretocmd{\todo}{\@bsphack}{}{}
\xapptocmd{\todo}{\@esphack}{}{}
\makeatother

\usepackage{booktabs}
\usepackage{csquotes}

\usepackage{graphicx} 
\usepackage{appendix}

\def\beq{\begin{equation}}
\def\eeq#1{\label{#1}\end{equation}}
\def\eeqn{\end{equation}}

\newenvironment{Eqnarray}%
   {\arraycolsep 0.14em\begin{eqnarray}}{\end{eqnarray}}
\def\beqa{\begin{Eqnarray}}
\def\eeqa#1{\label{#1}\end{Eqnarray}}
\def\eeqan{\end{Eqnarray}}

\let\bar=\overbar

\def\lsim{\mathrel{\raise.3ex\hbox{$<$\kern-.75em\lower1ex\hbox{$\sim$}}}}
\def\gsim{\mathrel{\raise.3ex\hbox{$>$\kern-.75em\lower1ex\hbox{$\sim$}}}}

\def\del{\partial}
\def\Dslash{\not{\hbox{\kern-4pt $D$}}}
\def\dslash{\not{\hbox{\kern-2pt $\del$}}}

\def\Dlr{\mathrel{\raise1.5ex\hbox{$\leftrightarrow$\kern-1em\lower1.5ex\hbox{$D$}}}}

\def\Pl{{\mbox{\scriptsize Pl}}}

\def\msb{{\bar{\scriptsize M \kern -1pt S}}}

\def\drb{{\bar{\scriptsize D \kern -1pt R}}}

\DeclareCiteCommand{\citejournal}[\mkbibbrackets]
  {\usebibmacro{prenote}}
  {\usebibmacro{citeindex}%
   \printtext[bibhyperref]{\printfield{journaltitle}}%
   \iffieldundef{volume}
     {}%
     {\setunit{\addspace}%
     \printtext[bibhyperref]{\printfield{volume}}}%
   \setunit{\addspace}%
   \printtext[bibhyperref]{(\printdate)}%
   \iffieldundef{pages}
     {}
     {\setunit{\addspace}%
     \printtext[bibhyperref]{\printfield{pages}}%
     }%
     }
  {\multicitedelim}
  {\usebibmacro{postnote}}
  
\DeclareCiteCommand{\citesubmit}[\mkbibbrackets]
  {\usebibmacro{prenote}}
  {\usebibmacro{citeindex}%
   \printtext[bibhyperref]{\printfield{journaltitle}}%
   \setunit{\addspace}%
   \printtext[bibhyperref]{(\printdate)}}
  {\multicitedelim}
  {\usebibmacro{postnote}}
  
  \DeclareCiteCommand{\citeconf}[\mkbibbrackets]
  {\usebibmacro{prenote}}
  {\usebibmacro{citeindex}%
   \printtext[bibhyperref]{\printfield{howpublished}}%
   \setunit{\addspace}%
   \printtext[bibhyperref]{(\printdate)}}
  {\multicitedelim}
  {\usebibmacro{postnote}}

\title{Design Initiative for a 10 TeV pCM Wakefield Collider}

\nolicence 

\notitlestamp 

\date{\today}

\addauthor{\footnotesize Spencer Gessner$^{1}$, Tim Barklow, Mark Hogan, Alexander Knetsch, Nathan Majernik, Brendan O'Shea, Michael Peskin, Ariel Schwartzman, Doug Storey, Caterina Vernieri,\\ \emph{SLAC National Accelerator Laboratory}}

\addauthor{Jens Osterhoff$^{2}$, Carlo Benedetti, Eric Esarey, Cameron Geddes, Simon Knapen, Remi Lehe, Simone Pagan Griso, Angira Rastogi, Carl Schroeder,\\ \emph{Lawrence Berkeley National Laboratory}}

\addauthor{Oksana Chubenko, Xueying Lu, \emph{Northern Illinois University}}

\addauthor{Vera Cilento, Marlene Turner, \emph{CERN}}

\addauthor{Brigitte Cros, Francesco Massimo, \emph{LPGP CNRS, Universit\'{e} Paris Saclay}}

\addauthor{Mariastefania De Vido, \emph{STFC}}

\addauthor{Matthias Fuchs, \emph{Karlsruhe Institute of Technology}}

\addauthor{Almantas Galvanauskas, \emph{University of Michigan}}

\addauthor{Joe Grames, Andrei Seryi, \emph{Jefferson Laboratory}}

\addauthor{Lindsey Gray, \emph{Fermilab}}

\addauthor{Thomas Grismayer, \emph{University of Lisbon}}

\addauthor{Nicole Hartman, \emph{Technical University Munich}}

\addauthor{Chunguang Jing, Rachel Margraf-O'Neal, Philippe Piot, \emph{Argonne National Laboratory}}

\addauthor{Sid Karkare, \emph{Arizona State University}}

\addauthor{Leily Kiani, \emph{Lawrence Livermore National Laboratory}}

\addauthor{Carl A. Lindstr{\o}m, \emph{University of Oslo}}

\addauthor{Michael Litos, \emph{University of Colorado Boulder}}

\addauthor{Patric Muggli, \emph{Max Planck Institute for Physics}}

\addauthor{Max Swiatlowski, \emph{TRIUMF}}

\addauthor{Robert Szafron, \emph{Brookhaven National Laboratory}}

\addauthor{Maxence Thevenet, \emph{DESY}}

\addauthor{Livio Verra, \emph{INFN}}

\addauthor{Eleni Vryonidou, \emph{Universtity of Manchester}}

\addauthor{
\begin{flushright}
{$^{1}$}sgess@slac.stanford.edu\\
{$^{2}$}josterhoff@lbl.gov
\end{flushright}
}

\abstract{\noindent
The community-driven Design Study for a 10 TeV pCM Wakefield Accelerator Collider introduced by this document is motivated by the 2020 ESPP Report emphasizing the need for advanced accelerator R\&D for future colliders, and the 2023 P5 Report calling for the ``delivery of an end-to-end design concept, including cost scales, with self-consistent parameters throughout" targeting the energy frontier.  This Design Study leverages recent experimental and theoretical progress from a global R\&D program with the goal of delivering a unified, 10 TeV Wakefield Collider concept. Wakefield accelerators provide ultra-high accelerating gradients which enables an upgrade path to extend the physics reach of a Higgs factory linear collider beyond the electroweak scale. Here, we describe the organization of the Design Study including timeline and deliverables, and detail requirements and challenges on the path to a 10 TeV Wakefield Collider.

}

\DeclareTOCStyleEntry[beforeskip=.2cm]{section}{section}

\begin{document}

\titlepage

\pagenumbering{arabic}\setcounter{page}{1}


\section{Executive Summary} 
The 2020 European Strategy for Particle Physics highlighted the importance of innovative accelerator technologies and recommended intensifying R\&D activities on advanced accelerators {\em (ESPP Rec. 3b)}~\cite{ESPP:2020}. The 2023 US Particle Physics Project Prioritization Panel (P5) Report~\cite{P5:2023} recognized the need for ``a 10 TeV parton center-of-momentum (pCM) collider to search for direct evidence . . . of new physics''. Since no technology is ready for building a 10 TeV pCM machine today, P5 calls for ``vigorous R\&D toward a cost-effective 10 TeV pCM collider based on proton, muon, or possible wakefield technologies'' {\em (P5 Rec. 4a)}. The 10 TeV Wakefield Collider Design Study has been formed in response to the 2023 P5 Report to provide a concept for an affordable, high-energy lepton collider that will enable discovery science at the energy frontier.

The Large Hadron Collider (LHC) is a monumental scientific achievement, most notably confirming the existence of the Higgs boson in 2012 and providing crucial experimental validation of the Standard Model of particle physics. However, many questions in particle physics and cosmology remain unanswered, including the nature of dark matter, the matter-antimatter asymmetry in the universe, and the hierarchy problem. Higher-energy colliders will offer the possibility to discover new physics and address these open questions, but we do not yet know the collision energy needed for new discoveries. LCVision~\cite{LCVisions:2025} proposes a Linear Collider Facility (LCF) as the next Higgs factory and electroweak-scale machine. Such a machine could uncover new physics through precision measurements, and those measurements would indicate the energy scale of new physics. High-gradient wakefield accelerators offer a unique opportunity to upgrade an LCF to directly probe the energy scale of new physics within the footprint of the existing facility. The 10 TeV Wakefield Collider Design Study will deliver an end-to-end concept for a very high-energy lepton collider, examining both the green-field case and the LCF upgrade case. While 10 TeV collision energy is already an ambitious target, we note that the wakefield collider concept is flexible and can be readily extended to reach higher energies, if motivated by precision electroweak measurements.

The 10 TeV Wakefield Collider Design Study is led by a global community of experts---particle theorists, HEP experimentalists, and accelerator scientists from national laboratories and universities---and is driven by a growing number of worldwide contributors who are motivated by the promise of a 10 TeV Wakefield Collider. The Design Study exists within a larger framework of initiatives pursuing future colliders. The 10 TeV Wakefield Collider is included as an energy frontier upgrade path for an LCF, as detailed in the LCVision document~\cite{LCVisions:2025}, and the 10 TeV Design Study operates in close collaboration with the ALEGRO~\cite{ALEGRO:2019}, HALHF~\cite{HALHF_ESPPU}, ALIVE~\cite{ALIVE_ESPP}, and XCC~\cite{XCC:2023} design studies. The physics case for the 10 TeV Wakefield Collider overlaps with the Muon Collider~\cite{MC:2024}, and commonalities are emphasized. The Design Study participants have begun meeting on a regular basis, both online and in-person at the recent ALEGRO Workshop~\cite{ALEGRO2025}.

The Design Study team has started to examine the physics cases for $e^+e^-$, $e^-e^-$, and $\gamma\gamma$ collisions at 10 TeV pCM. 
The design of the interaction point (IP) and detector follows from the analysis of the physics case for each of the collision scenarios. The Design Study includes working groups that address the individual components of the collider within a self-consistent framework. The Systems and Optimization Working Group will evaluate collider design choices with the goal of minimizing cost and environmental impact while maximizing the physics reach of the collider.

The deliverables for the 10 TeV Wakefield Collider Design Study include: 1) a description of the physics program for 10 TeV $e^+e^-$, $e^-e^-$, and/or $\gamma\gamma$ collisions, 2) an end-to-end design concept, including cost scales, with self-consistent parameters, and 3) roadmaps and resource estimates for medium-term R\&D and required demonstrator facilities. Although the Design Study is in its early stages, we have already identified innovative methods and technologies that will also benefit near-term Higgs factories. The 10 TeV Wakefield Collider Design Study requires dedicated funding in order to carry out the detailed work that is needed for an accurate costing of the machine. Investment from the European High Energy Physics community will help the Design Study to perform this work and return critical input to the community as it evaluates the most affordable paths toward 10 TeV pCM collisions.

\newpage

\section{Introduction} 
The 10 TeV pCM Wakefield Collider Design Study forges an affordable path for linear colliders to reach the energy frontier~\cite{LCVisions:2025}. The Design Study is formed in response to the 2023 US Particle Physics Project Prioritization Panel (P5) Report~\cite{P5:2023} which calls for ''the delivery of an end-to-end design concept, including cost scales, with self-consistent parameters throughout.'' Both the 2020 European Strategy for Particle Physics~\cite{ESPP:2020} and the 2023 P5 Report recognized the importance of fundamental accelerator R\&D and highlighted wakefield accelerators as possible technology for future particle physics applications. 

R\&D on wakefield accelerators is currently being pursued at research laboratories around the world~\cite{Power:2022,Muggli2022}. The last two decades have seen exceptional progress in wakefield accelerator science, with year-over-year increases in the number of publications from the field~\cite{SnowmassAF6}. Accelerator science is critical to the R\&D process, but the next leap for high-gradient wakefield accelerators requires breaking out of the lab and into society. Wakefield accelerators will have broad applications across scientific disciplines, as well as in industry and medicine, owing to their compact nature. Wakefield accelerators have demonstrated accelerating gradients from 100 MV/m to 100 GV/m, far in excess of what is typically achieved with radiofrequency technology. Advances in wakefield accelerator capabilities are the result of improvements in beam test facilities which are able to provide intense particle and laser beams with spatio-temporal control over their profiles~\cite{Power:2022,Picksley2024,Shrock2024}. The coming decade will see a transition from proof-of-principle experiments to first applications, such as free-electron lasers~\cite{Galletti2024}, that will not only enable new science, but also demonstrate robust and reliable operation of wakefield accelerator systems.

The 10 TeV Wakefield Collider Design Study channels our growing understanding of wakefield technology towards a future collider application. The Design Study benefits from a network of related efforts that seek new paths toward high energy collisions. LCVision proposes a Linear Collider Higgs Factory as the next global collider, and includes wakefield accelerator technology as part of its upgrade path~\cite{LCVisions:2025}. The HALHF design study~\cite{Foster2025} considers the staging of beam-driven plasma accelerators at Higgs Factory energies, which directly feeds into the work of the 10 TeV Wakefield Collider Design Study. In addition, the HALHF collaboration has developed a system code for collider optimization and costing which will be incorporated into our Design Study. The XCC~\cite{XCC:2023} proposes a new approach for $\gamma\gamma$ collisions for a future Higgs Factory, and some of XCC's innovations have been adapted for the 10 TeV case~\cite{Barklow2023}. The physics case for a 10 TeV lepton collider has been detailed by design studies for the Muon Collider~\cite{MC:2024}, and we extend this work to the Wakefield Collider case where applicable.


As wakefield technology has matured, the path toward 10 TeV collisions has become clearer. So too have the challenges. Wakefield colliders will require precision alignment and stability, beyond what has been considered for the most demanding linear collider concepts, such as CLIC~\cite{CLIC:2012}. Plasma wakefield accelerators bring two unique challenges: staging of plasma modules~\cite{Lindstrom:2020} and positron acceleration~\cite{Cao2024positrons}. In order to maximize the luminosity-per-power of a wakefield collider, the beams must be focused to sub-nanometer spot sizes. The fields experienced by the colliding particles are enormous, leading to extreme beamstrahlung and broad luminosity spectra~\cite{Barklow2023}. For these reasons, the 10 TeV Wakefield Collider Design Study is investigating $e^+e^-$, $e^-e^-$, and $\gamma\gamma$ collisions as possible operational scenarios.

The Design Study examines trade-offs between the different collision scenarios. $e^-e^-$ and $\gamma\gamma$ collisions avoid the challenge of positron acceleration in plasma, but do not necessarily deliver a complete physics program at 10 TeV. In parallel with our effort to understand the physics reach of the machine, we analyze beam-beam effects and beam-induced-backgrounds at the interaction point (IP) which informs the detector design and the Machine-Detector Interface (MDI). We then work outwards to define the Beam Delivery System (BDS) and main linacs based on beam-driven plasma, laser-driven plasma, and structure wakefield accelerators. The colliding beams must have very low emittance and new types of particle sources are under investigation. 
A dedicated Systems Integration Working Group is tasked with an overall optimization of the design while ensuring self-consistency. We aim to deliver a design that describes the necessary infrastructure and hardware with sufficient detail to enable a rough cost estimate of the machine and its operation.

This document articulates the physics case for a 10 TeV Wakefield Collider, and it discusses the design of the facility starting from the IP and working outward along the accelerator. We highlight synergies between our Design Study and other efforts, including methods and technologies that will benefit Higgs factories in the near-term. We conclude with a description of the Design Study timeline and deliverables.

\section{Physics at 10 TeV Parton-Center-of-Momentum}
\subsection{Why 10 TeV?}

A 10 TeV parton center-of-momentum collider would represent a major leap forward in our understanding of fundamental physics, probing energy scales well beyond those accessible at the Large Hadron Collider (LHC). One of its central motivations is to explore the nature of electroweak symmetry breaking and the Higgs sector: while the discovery of the Higgs boson confirmed the mechanism of mass generation, many fundamental questions remain unresolved. A higher-energy collider could test whether the Higgs boson is elementary or composite, reveal possible strong dynamics in the Higgs sector, or uncover the existence of additional Higgs bosons. Precision measurements of the Higgs self-couplings and interactions with other particles would offer sensitivity to deviations from Standard Model predictions and potential insights into the baryon asymmetry of the universe. Simultaneously, a 10 TeV collider would significantly extend the reach for direct searches for physics beyond-the-Standard Model (BSM), including heavy superpartners predicted by supersymmetry, dark matter candidates such as WIMPs, new gauge bosons, extra dimensions, or provide evidence of fermion substructure. 
Analogously to the extensively studied Compact Linear Collider (CLIC) \cite{Roloff:2018dqu} and muon collider option \cite{AlAli:2021let, Aime:2022flm,Accettura:2023ked}, these goals motivate a detailed exploration of different initial-state configurations for wakefield colliders including $e^+ e^-$, $e^-e^-$, and $\gamma \gamma$ collisions,  to fully exploit the physics potential across a broad range of processes and exotic signatures.

Each of these collider configurations offers unique capabilities that complement one another in probing new physics, as will be detailed in the design study.
Broadly speaking, all three options can discover new heavy particles, in particular if they are charged under the electroweak force. 
The presence of such electroweak particles below 10 TeV is a generic expectation of models of electroweak symmetry breaking and weakly interacting particles feature also as dark matter candidates.
%
%
In addition to direct searches for new particles, a 10 TeV collider would enable precise tests of the Standard Model at the highest energies ever conducted by measuring electroweak parameters, flavor observables, and CP violation in a regime that is complementary with high intensity, low-energy observables.
The collider would also provide new insights into the top quark, the heaviest known elementary particle, which could play a crucial role in electroweak symmetry breaking. High-energy collisions could test whether the top quark has substructure or interacts via new forces, while also shedding light on the strong interaction (QCD) at extreme energies. Precision studies of jet structures, potential new bound states, and saturation effects in QCD could further refine our understanding of fundamental forces.


Dedicated studies have been initiated to characterize the distinct physics potential of a 10 TeV lepton collider. While the physics motivations will be the same as those explored for the 10 TeV muon collider, we expect qualitative differences due to the importance of beam-beam interactions for a linear wakefield collider.
These studies will establish the minimal required luminosity  for each beam configuration to achieve the above-mentioned physics goals.
This, in turn, will be input for the various accelerator design working groups.

\subsection{Physics at 10 TeV Wakefield Collider}

A wakefield collider differs significantly from an equivalent pCM muon collider due to the unique beam--beam interactions that arise from ultra-relativistic, high-charge beams. One of the key differences is the beamstrahlung, which generates large numbers of highly-energetic secondary particles, including electrons, photons, and even positrons. Although some of these secondary particles could contribute to signal events, they can also introduce backgrounds that must be carefully managed. Understanding and managing these effects is crucial for optimizing the physics performance of a wakefield-based collider.  

Another fundamental difference lies in the acceleration mechanism itself, which is inherently asymmetric if using plasma as a medium. Accelerating positrons in a plasma-wakefield accelerator is significantly more challenging than accelerating electrons, posing a technical hurdle for realizing an electron--positron (\(e^+e^-\)) collider. Given this constraint, it is important to explore alternative configurations such as electron--electron (\(e^-e^-\)) and photon--photon (\(\gamma\gamma\)) colliders. These options could potentially be realized on a shorter timescale and with higher luminosity, making them attractive alternatives to the baseline \(e^+e^-\) design. Considering such variations in collider design could be key to leveraging the advantages of wakefield acceleration while addressing its unique challenges.

Beamstrahlung may present opportunities in the context of a wakefield collider: Standard Model and beyond-the-Standard-Model processes involving particles well below the maximum collider energy often have cross sections that scale as \(1/E_{COM}^2\). This implies that relatively low-energy collisions, originating from particles produced via beamstrahlung, can, in some cases, dominate the overall production rate. Rather than being merely a nuisance, beamstrahlung could thus play an active role in enhancing certain physics signals, potentially reshaping expectations for collider phenomenology.
This is particularly relevant when considering round vs.~flat beam profiles.

For an electron--electron (\(e^-e^-\)) collider, the absence of Drell-Yan production for new particles that do not carry lepton number presents a limitation. While some of this deficit can be compensated by \(ZZ\) fusion, an intriguing possibility arises from beamstrahlung-induced positrons. These positrons can enable Drell-Yan production channels that would otherwise be suppressed in a pure \(e^-e^-\) environment. In certain scenarios, these processes may contribute significantly to new physics searches. A careful quantification of these trade-offs across a range of Standard Model and beyond-the-Standard-Model processes will be essential in determining the optimal collider configuration. Such studies will ultimately help guide decisions on which collision flavors and beam profiles should be pursued with targeted R\&D efforts.




\section{Detector}
\subsection{Machine-Detector Interface}
For the 10 TeV environment at an advanced $e^+e^-$ or $\gamma \gamma$ collider, we are studying the beam-induced backgrounds (BIB, described in Sect.~\ref{subsec:beam-beam-int}) and their interplay on the detector design. The expected predominant backgrounds in our detector are: the $e^+e^-$ pairs produced by beamstrahlung and the photon-initiated production of hadrons.

The deliverables for our group will include: (1) the distribution of the BIB from the interaction region (assuming a uniform magnetic field), (2) the occupancy in the innermost layers of the detector, and (3) the arrival time of the particles hitting the innermost detector layers. To develop these studies quickly, we leverage the recent innovations for studying the BIB with the $C^3$~\cite{Ntounis2024} and Muon Collider detector designs~\cite{Collamati2021}. 
These workflows leverage the \texttt{KEY4HEP}~\cite{Key4hep:2023rka} software stack (future colliders software setup that ships all the key ``physics and detector optimization'' packages),
and we use \texttt{DDSim} to simulate the interaction with the detector with \texttt{GEANT4}~\cite{Agostinelli2003}.
As part of this design study, we plan to synergize our efforts with the developments from the Beam-Beam Interactions team, including testing their novel particle-in-cell codes for our detector design metrics.

\subsection{Detector}
The key design parameters for an effective detector depend heavily on the beam type ($e^+e^-$, $e^-e^-$ or $\gamma \gamma$), beam configuration (round vs. flat), the machine-detector interface (MDI), the beam delivery system (BDS), and the targeted physics goals. A careful evaluation of these factors is essential to identifying a cost-effective and scalable path for a future collider.

Leveraging the \texttt{KEY4HEP} software stack and existing detector models as a starting point, it is expected to modify the detector layout to fit the 10 TeV environment and expected dynamic range of objects' properties for this environments and various collision types. After an initial re-dimension of the main detector elements, a more careful re-optimization of the layout and, in parallel, a study of the environment expected will lead to writing the requirements for each sub-system and identify technologies that could fit those requirements and R\&D priorities.

Among the possible beam configurations, $e^+e^-$ collisions provide the most straightforward way to access the largest variety of physics. In particular, they are ideal for Higgs precision measurements and beyond-standard-model (BSM) searches. However, they also present significant challenges due to intense beamstrahlung radiation, which produces a substantial flux of secondary high-energy electrons, positrons, and photons. Additionally, the current difficulties in positron acceleration raise the need for alternative strategies.  One approach is
asymmetric collisions,  in which the positron beam has lower energy and density, while the electron number density or repetition rate is increased to sustain luminosity. Another possibility is leveraging beamstrahlung-generated positrons in $e^-e^-$ collisions, potentially enabling a faster path to a working collider. These considerations necessitate a careful detector design that effectively mitigates beam-beam backgrounds without overcompensating.

The BDS layout and its distance from the interaction point impose fundamental constraints on luminosity, detector acceptance, and calorimeter depth for high-energy forward particles. Understanding the bunch structure whether single bunch collisions or bunch trains with specific spacings is crucial, as is assessing the impact of the BDS positioning on luminosity. Additionally, backscattered electrons from the MDI introduce timing challenges for the innermost vertex and tracking detectors. To distinguish signal from background, at least nanosecond-level time resolution is required. Fortunately, state-of-the-art silicon detector technology enables such precision for charged-particle measurements.

For $\gamma\gamma$ collisions, the detector design must accommodate the complexities of beam optics and the MDI, particularly addressing backscattering from Compton interaction points and the primary interaction point. The final focusing system near the interaction region plays a crucial role in ensuring precision. Some studies suggest replacing conventional final focusing magnets with plasma lenses to improve beam control. However, integrating plasma lenses into the detector environment demands careful consideration of key factors such as positional stability, dimensional constraints, power requirements, heat dissipation, and cooling mechanisms to maintain optimal performance.

\section{Accelerator}
We describe the components of the 10 TeV Wakefield Collider starting from the interaction point (IP) and working outward to the particle sources. Each component is assigned to a Working Group. This section concludes with a description of the Systems Integration and Optimization Working Groups that are tasked with a holistic assessment of the collider.

\subsection{Beam-Beam Interactions}
\label{subsec:beam-beam-int}
The Beam-Beam Working Group will address challenges related to the collisions of extremely high-energy and extremely dense particle beams. When high-energy beams collide, they undergo beamstrahlung and radiate high-energy photons~\cite{YokoyaChen:1989}. The beamstrahlung process depletes the energy of the colliding beams and leads to beam-induced backgrounds as beamstrahlung photons convert to electron-positron pairs in the field of the opposing beam. The efficacy of a collider is measured by its luminosity-per-power ($\mathcal{L}/P$)~\cite{Roser2023}. In cases with large beamstrahlung, we must also consider the luminosity spectrum of the collisions. In previous work, we counted the luminosity within 20\% of the collider center-of-mass ($\mathcal{L}_{20}$) as contributing to the physics reach of the collider, and used the metric $\mathcal{L}_{20}/P$ to quantify the collider efficacy~\cite{Barklow2023}. Our ongoing studies have shown that this metric might be too simple to fully capture the physics at 10 TeV. In order to enhance the physics reach of a 10 TeV pCM Wakefield Collider, we must maximize the luminosity while minimizing the power needed to produce the colliding beams. We consider collisions between electrons and positrons, electrons and electrons, and photons and photons as different avenues to maximize $\mathcal{L}/P$.

The deliverables for the Beam-Beam Working Group include a comparison of different collision types ($e^+e^-$, $e^-e^-$, and $\gamma\gamma$) taking into account the full $\mathcal{L}/dE$ spectrum. For each of these collision types, we will determine optimal beam parameters, and we will also estimate the beam-induced background in each scenario. We will develop new theoretical frameworks for analyzing cases with extreme beamstrahlung. Particle-in-Cell (PIC) codes are the principle method for modeling Beam-Beam interactions. We will utilize the legacy codes GUINEA-PIG and CAIN to benchmark new High-Performance Computing (HPC) codes including WarpX, OSIRIS, and VLPL. The HPC PIC codes will include the physics of the legacy codes, and deploy new models to cover extreme beamstrahlung regimes. The HPC PIC codes are critical for the design of the 10 TeV pCM Wakefield Collider, but they will also broadly benefit the HEP field as replacements for legacy codes that are currently being used to model the next generation of Higgs Factories. We will deliver fast, high-resolution, user-friendly PIC codes that will become the community standard for Beam-Beam modeling for all future HEP colliders.

\subsection{Beam Delivery System}
The Beam Delivery System (BDS) design for a 10 TeV wakefield collider is still in its early stages, with current efforts focusing on scaling and adapting existing concepts. As a starting point, we consider the CLIC BDS designs at 380 GeV, 3 TeV, and 7 TeV, estimating that a 10 TeV BDS would require an approximately 8 km beamline, assuming flat beams and an interaction point (IP) with L$^*$ = 6 m. Key challenges include synchrotron radiation in bending magnets, the feasibility of a triplet final focus system for round beams instead of a doublet, and addressing chromaticity correction issues with large emittances and small beta functions, potentially necessitating a traditional Final Focus System (FFS) approach. Several techniques and technologies are under consideration, including laser-based energy collimation, plasma lenses, and nonlinear collimation schemes inspired by CLIC. Questions of stability, aberrations, complexity, and limitations~\cite{Lindstrom2018} will also be studied for active and passive plasma lens concepts. 
Additionally, strategic decisions regarding the number of IPs-whether to feature one or two-will significantly influence the final layout and optimization of the BDS. Emphasizing collaborative efforts, the working group will regularly discuss common problems and potential common solutions, using the 7 TeV CLIC BDS as the foundational reference~\cite{Manosperti2023}. Through sustained collaboration and iterative refinements, a robust and efficient BDS+IP design for the 10 TeV wakefield collider will be developed.

\subsection{Main Linac}
The key advantage of wakefield accelerator technology is that it can be leveraged to accelerate lepton beams with 1-100 GeV/m accelerating gradient, as demonstrated in recent experiments~\cite{Corde2016}. However, these large accelerating gradients are only sustained for short distances before the drive beam is depleted of energy. Beam and laser-driven plasma accelerators must be staged together, and in this case the geometric gradient defined as the total energy gained over the length of the linac must be maximized. LC Visions proposes 11 km-long linac arms for an ILC-type machine~\cite{LCVisions:2025}, which implies a minimal geometric gradient of 500 MeV/m assuming some room for overhead. Note that the structure wakefield accelerators do not have the same staging challenges as plasma-based accelerators, and in that case the in-structure gradient must be maximized. We describe challenges and plans for laser-driven plasma (LWFA), structure-based wakefield (SWFA), and beam-driven plasma (PWFA) in the following sections.

\subsubsection{LWFA Linac}
The LWFA Linac Working Group will address the design challenges of an LWFA-based linac for accelerating electron and positron beams to 5 TeV energies. LWFAs sustain accelerating gradients of 10-100 GV/m, up to 1000 times higher than conventional RF technology \cite{Esarey2009}, making them promising for a cost-effective future collider. An LWFA linac relies on staged, independently powered LWFA sections, with plasma mirrors coupling laser drivers in and out over short distances. This approach enables high beam energies while maintaining high accelerating gradients without increasing laser energy. Additionally, LWFAs produce ultrashort particle bunches, reducing beamstrahlung and overall power consumption for a given luminosity \cite{Schroeder2022}.

A conceptual design for an LWFA-based linac for collider applications has not been completed, but preliminary studies have identified key laser and plasma parameters \cite{Schroeder2010, LWFA_Schroeder_NIMA_2016, Schroeder_2023}. For instance, LWFA properties depend on the choice of the operational plasma density, which sets the energy gain provided by the stage, its length, the characteristic charge that can be accelerated, the laser energy required to drive the stage, etc. Considerations concerning minimization of the total accelerator length, wall-plug power, and beamstrahlung effects, while reaching the desired luminosity (e.g., $>10^{34}$ cm$^{-2}$ s$^{-1}$ for >1 TeV center of mass energies) suggest operating at a density on the order of $\sim 10^{17}$ cm$^{-3}$ \cite{Schroeder2010,Schroeder2022}. This requires laser drivers with an energy of 10s of J operating with a repetition rate of 10s of kHz. Each LWFA stage will then provide multi-GeV energy gains to beams with a charge of 100s of pC, over a distance of 10s of cm. Reaching TeV-class energies necessitates cascading 100s of stages, including their interfaces. High efficiencies (several 10s of percent) from wall-plug to laser driver production, from driver to wakefield excitation, and from wakefield to beam are required. In order to ensure that the particle beam can be effectively focused at the interaction point and to enable staging, high beam quality (i.e., relative energy spreads $< 1\%$   and emittances <100 nm) must be maintained throughout the accelerator. This implies that the chosen acceleration regime must be free from instabilities which could lead to emittance degradation and/or beam loss. In addition, each compact inter-stage coupling, e.g. provided by plasma mirrors, must be carefully designed to ensure the efficient deflection of the laser driver from the previous stage and the focusing of the laser driver for the next stage. The energetic particle beams must pass through the interface and be injected into the next stage without quality degradation due to interactions with the mirror \cite{Zingale2021} or with the deflected laser pulse \cite{Streeter2020}. 

The deliverables for this Working Group include the design of LWFA stages providing high-quality, high-energy, and high-efficiency acceleration of (polarized) electrons and positrons beams for a given laser technology (i.e., fixed laser energy and wavelength). Example of metrics to assess the performance of a given design include evaluation of maturity of a given technology, minimization of linac footprint, and maximization of particles delivered to the BDS for fixed wallplug power. The tools used for this activity are theoretical LWFA scaling laws \cite{Schroeder2010,Schroeder2022}, reduced physics models, and fully self-consistent particle-in-cell simulations.

\subsubsection{SWFA Linac}
Structure-based wakefield acceleration (SWFA) linacs offer the advantage of identical acceleration for electrons and positrons but lower gradients compared to their plasma-based counterparts~\cite{Jing2022, lu2022, shao_ipac2022,peng_ipac2019,picard-2022,freemire,merenich}. To better integrate SWFA into the 10 TeV wakefield collider design, the SWFA Linac Working Group will first explore the fundamental limits of SWFA linacs, including the maximum achievable accelerating gradient while maintaining beam quality, mitigating beam breakup and other instabilities, and efficiently coupling power into the accelerating structures. A critical challenge for a multi-TeV collider is staging: efficiently transferring beams between sequential accelerator modules while minimizing emittance growth and preserving synchronization. We will also refine linac parameters to best match the beam configurations defined by other working groups. A crucial metric for assessing collider feasibility is the luminosity-to-power ratio, which is directly influenced by acceleration efficiency, beam emittance preservation, and power-to-beam transfer efficiency. Additional key performance indicators include accelerating gradient, energy spread control, and the overall scalability of the acceleration scheme to collider-relevant parameters. To address these challenges, the working group will investigate novel structure geometries, advanced materials for high-gradient operation, innovative drive-beam configurations, and staging strategies to maximize energy extraction and beam stability.

The deliverables for the SWFA Linac Working Group will include a comparative study of structure designs and materials capable of sustaining high gradients while minimizing breakdown rates. We will develop and validate advanced numerical models to predict beam dynamics, wakefield effects, and staging challenges in SWFA structures. Additionally, we will evaluate the optimal beam loading strategies to optimize energy transfer efficiency and stability. These efforts will contribute directly to the conceptual design of a 10 TeV pCM Wakefield Collider, providing a clear roadmap for SWFA-based linac development. Beyond collider applications, the working group's results will benefit the broader HEP community by advancing compact, high-gradient accelerator technologies for applications in future lepton colliders~\cite{Jing2022}, advanced light sources~\cite{Piot2023,MargrafONeal2024,Zholents:2025ahe}, and high-brightness beam-driven experiments.

\subsubsection{PWFA Linac}
Beam-driven plasma wakefield accelerators (PWFAs)~\cite{chen1985acceleration} transfer energy from wake-exciting particle bunches into a trailing witness bunch with accelerating gradients on the order of 1-100 GV/m~\cite{blumenfeld2007energy}.
Two main drive bunch species could be used: protons and electrons.
In the case of protons, the large amount of energy contained within a single bunch produced by synchrotrons (e.g., $20\,$kJ from Super Proton Synchrotron at CERN) allows for wakefield generation and acceleration to high energies in a single long plasma stage\,\cite{caldwell2009proton}. 
Such proton bunches are longer than the typical plasma wavelength, necessitating self-modulation instability to create microbunches that drive large-amplitude wakefields\,\cite{adli2018acceleration}. 
Short proton bunches could overcome this challenge~\cite{Farmer_2024}, but requires significant R\&D.
In the case of electrons, the drive-beam energy is normally on the order of tens of GeVs. 
This means that the energy gain of the witness bunch in a single stage is limited to a similar amount. 
Hence, to reach TeV energies, a sequence of stages driven by independent drivers will be necessary. 
Staging is therefore the most outstanding challenge to realize an electron-driven PWFA linac. 
In particular, the main challenge is the transport of the witness bunch from one stage to the following, while preserving the beam quality and maintaining a high geometric accelerating gradient. 

While the acceleration of positrons by PWFA has been previously demonstrated~\cite{Corde2015positrons}, it remains a challenge to do so while maintaining beam quality due to the intrinsic charge asymmetry in plasmas~\cite{Cao2024positrons}. 
Similarly, the acceleration of flat beams by PWFA may also face challenges on maintaining beam quality, due to coupled betatron oscillations in the transverse planes resulting in emittance mixing, however, several techniques for overcoming this limitation have been proposed~\cite{Diederichs2024FlatBeams}.

Inspired by progress on beam quality in electron-beam-driven PWFAs,  such as preserved emittance \cite{lindstrom2024emittance}, energy spread \,\cite{joshi2018plasma,pompili2021energy,lindstrom2021energy} in PWFAs, and lasing of the accelerated bunch, the research community is already working on Higgs-Factory designs\,\cite{Foster2023,cros2017towards} with c.o.m. energies up to 550 GeV.

The PWFA linac working group aims to go beyond and evaluate the suitability of PWFAs to cost- and space-effectively accelerate collider-quality electron beams into the TeV energy range.
Physics constraints and design challenges of multi-stage PWFAs \,\cite{lindstrom2021self,PhysRevLett.131.135001} will be explored in the different energy regimes from few GeVs to several TeV.
Finally, results drawn from the study will inform overall LINAC design and facilitate optimization of program cost and facility footprint.

\subsection{Drivers}

\subsubsection{Laser Drivers}
The Laser Drivers Working Group will address challenges related to the development of high peak and average power lasers for driving the laser wakefield accelerators. Over the last decade the LWFA community provided multiple iterations on key anticipated laser driver performance parameters, the latest of which is summarized in the 2023 Report of the Basic Research Needs Workshop on Laser Technology co-sponsored by DOE, NSF, and DOD \cite{Kling2024}. Later iterations further refined the target operational parameter space for laser drivers. For laser drivers operating at around 1~$\mu$m, these include a pulse energy of 10~J at 50~kHz and of 50~J at around 1~kHz to target quasi-linear and Bubble regimes, respectively. For 1~$\mu$m lasers, pulse duration will need to be in the region of 70-80~fs. For other wavelengths, pulse duration requirements scale proportionally with wavelength, for example, 300-1000~fs at 10~$\mu$m. Target wall-plug efficiency is > 20\% in the near-term with a > 30\% requirement in the long-term. Pulse and beam quality requirements as currently specified by the LWFA community include a pre-pulse contrast better than 10$^3$ for a ns-ps window, a beam Strehl ratio $>$0.95, with a high degree of beam symmetry, a pulse energy stability better than 1\% root-mean square (rms) and a beam pointing stability better than 0.1~$\mu$rad rms. The main technological challenge is to achieve this very high average power scaling into the 50~kW - 500~kW range with 100-300~TW peak power ultrashort pulses, which is orders of magnitude beyond the current state-of-the-art in ultrafast lasers. Over the same decade laser technological pathways to achieve such extreme power and energy scaling have been introduced based either on coherently combined fiber lasers, or new solid-state laser materials (e.g. Tm:YLF or Yb:YAG) that can sustain much higher powers than the currently used Ti:sapphire ``workhorse" lasers. Meanwhile, high-pressure $CO_2$ gas amplifiers show potential for efficient operation at around 10~$\mu$m. However, all these pathways require significant further development.

The deliverables for this Working Group will include iterative refinements of laser driver parameters specifically required for the envisioned 10 TeV LWFA collider by close collaboration with the LWFA community, and further iterations on laser technology pathways to meet such requirements by the Working Group. Technological routes under investigation will include the two solutions based on new power scalable direct CPA schemes with solid-state media and coherent combination of fiber lasers, but also optical parametric chirped pulsed amplification (OPCPA), thin-disk, and titanium-doped sapphire pumped by diode-pumped solid-state lasers (DPSSLs), and other approaches. The Laser Drivers Working Group will also discuss the development of key enabling technologies for laser drivers, including new fibers, large-aperture gain materials, diffraction gratings, high power diode lasers, thermal management schemes, high resilience optics and electronics and control systems. The main benefits and challenges of each technological pathway will be discussed and analyzed, considering the current state-of-the-art and projected future developments. Findings will be distilled into a set of recommendations on the most promising research and development avenues and a roadmap for the realization of LWFA drivers.

\subsubsection{Beam Drivers}
In PWFAs and Collinear Wakefield Acceleration (CWA) SWFAs, drive beams with charge ${\cal O}(10\text{ nC})$ are typically required with duration commensurate with the excitation of high-amplitude terahertz-frequency wakefields~\cite{Lemery:2022uxo}. The infrastructure needed to form, accelerate, and manipulate these drive bunches represents a significant portion of the overall accelerator cost.
Optimizing the parameters of the drive beam and tailoring its phase space are critical strategies for enhancing efficiency~\cite{Tan2022}. For instance, carefully tailoring the drive-beam current profile enables transformer ratios $>2$, while properly controlling the longitudinal phase-space correlation can increase the effective interaction length and help mitigate beam instabilities such as single-bunch beam break-up in SWFAs or hose instability in PWFAs, thereby improving energy transfer efficiency.
In addition to the CWA configuration common to PWFA and SWFA, SWFAs can also be implemented using a two-beam acceleration (TBA) configuration similar to the CLIC design. In this latter case, the drive bunches are usually organized as a bunch train with total charge ${\cal O}(1\text{$\mu$C})$ and periods at a sub-harmonic of the fundamental accelerating mode. The drive beam system for the PWFA and SWFA linac options must provide stable, high-charge beams to drive large amplitude wakefields. Tolerances for plasma-driven accelerators are extremely tight, so the jitter emittance of the drive beam linac must be minimized. Despite the apparent difference between the TBA and CWA drive beam requirement, the drive-beam complexes share notable similarities. In fact, in its most recent iteration, the HALHF design has adopted a CLIC-like drive-beam complex for PWFA, incorporating an L-band linac with combiner rings~\cite{Foster2025}.

This working group will explore the generation, acceleration, and manipulation of drive bunches to excite electromagnetic wakes in P/SWFA, building on recent studies by the HALHF collaboration. The focus will be on delivering stable drive beams to P/SWFA modules while investigating instability mitigation strategies and optimizing phase-space control to enhance efficiency.

\subsection{Particle Beam Source }
The selection of particle beam sources will ultimately be determined by the choice of collider, whether an electron-positron ($e^{+}e^{-}$), electron-electron ($e^{-}e^{-}$), or gamma-gamma ($\gamma \gamma$) collider. At the same time, the cost and performance of these sources will influence the final collider design decision. This working group will assess various established and emerging techniques for electron and positron generation, focusing on their achievable brightness and estimated cost. We note that 4D beam brightness is directly proportional to the luminosity-per-power of the machine, which is a key figure of merit for the collider. Once a collider choice is finalized, a detailed design and cost analysis for the corresponding particle sources will be conducted.

Photoinjectors, which generate electrons via photoemission in a strong accelerating electric field, are the preferred choice for high-brightness electron sources. Current TeV-scale electron accelerator designs, such as those for the International Linear Collider (ILC), rely on damping rings to reduce emittance and achieve high brightness while also ensuring the asymmetric transverse emittances required for $e^{+}e^{-}$ collisions \cite{rubin_2015}. However, recent research suggests that photoinjectors alone may be capable of achieving comparable or even superior brightness with the desired asymmetric emittances. If confirmed, this advancement could eliminate the need for damping rings in electron sources, significantly reducing both cost and the overall footprint of the accelerator.


Positron generation typically involves colliding a high-energy (multi-GeV) electron beam with a target, or alternatively by direct illumination with high-energy (multi-MeV) gamma-ray beams to produce positrons, followed by damping rings to achieve the required high brightness. The need for both a high-energy electron accelerator to produce positrons and damping rings to enhance their quality significantly increases the cost and spatial footprint of an $e^{+}e^{-}$ collider. These factors will play a crucial role in the decision between an $e^{-}e^{-}$ or $e^{+}e^{-}$ collider.

While alternative methods for positron production are under exploration, they remain in the early stages of development~\cite{Streeter2024}. Large-scale European facilities such as EuPRAXIA~\cite{Sarri2022} and EPAC in the UK are designing laser-driven positron beamlines for programmatic studies of positron acceleration in plasmas. This working group will assess their feasibility and potential impact on future collider designs.


High-brightness beams generated by particle injectors and particle sources based on wakefield accelerator concepts are promising sources of ultra-low emittance beams~\cite{Fuchs2022}. In particular, high-gradient accelerators can generate and rapidly accelerate particle beams to relativistic energies. The rapid acceleration and strong confining fields enhance beam brightness. Wakefield-based sources also allow for the generation of ultrashort electron bunches, which is beneficial for reaching high luminosities with high energy-efficiency.

\subsection{Systems Integration, Optimization, and Environmental Impact}
The P5 Report specifically highlights ``self-consistent parameters throughout" the design. Each of the Working Groups will develop their component designs subject to the self-consistency constraint, but Systems Integration Working Group is ultimately responsible for ensuring a unified design. The Systems Integration WG will assess the design for possible inconsistencies and seek to resolve them. The Systems Integration WG, along with the Environmental Impact WG, is responsible for the overall optimization of the design, where the key parameter to be optimized is the cost of the collider. The 10 TeV Wakefield Collider Design Study will work closely with the HALHF study and make use of their system code ABEL to optimize the design~\cite{Foster2025}.

The Environmental Impact Working Group aims to minimize the environmental footprint of the 10 TeV pCM Wakefield collider. To achieve this, environmental considerations will be integrated into the design process from the outset. Tasks of the working group include defining energy efficiency metrics, evaluating the collider design for optimal efficiency, and determining power requirements. Furthermore, the group will develop strategies to minimize environmental impacts during construction and  operation, by e.g. investigating the implementation of energy recycling systems or optimizing beam parameters at the collision point. The $C^3$ design study has developed a framework for the analysis~\cite{Breidenbach2023}.
The  deliverable of the Environmental Impact Working Group is an assessment of the environmental impacts associated with the collider construction, operation, and decommissioning.



\section{Conclusion: Synergies and Deliverables}
The 10 TeV Wakefield Collider Design Study must solve a number of challenges in order to create a viable path towards a future collider. The Design Study should innovate in order to address technical risks, and some of these innovations will benefit near-term machines. Although our study is still in its early stages, we have already identified several tasks which will benefit the design and optimization of a future Higgs factory. These include: 1) High-Performance Computing (HPC) Particle-in-Cell (PIC) codes for modeling beam-beam interactions, 2) Laser-based collimation systems~\cite{Zimmermann1999}, and 3) Non-intercepting diagnostics for ultra-intense beams.

The Design Study has natural synergies with any future linear collider Higgs factory as a long-term upgrade to such a machine. The upgrade path for a linear collider facility is detailed in the LCVision document~\cite{LCVisions:2025}. The 10 TeV Wakefield Collider Design Study both complements and benefits from on-going design efforts including ALEGRO~\cite{ALEGRO:2019}, HALHF~\cite{Foster2023}, ALIVE~\cite{Farmer_2024}, XCC~\cite{XCC:2023}, and the Muon Collider~\cite{MC:2024}.

The deliverables for the 10 TeV Wakefield Collider Design Study include: 1) a description of the Physics Program for 10 TeV $e^+e^-$, $e^-e^-$, and/or $\gamma\gamma$ collisions, 2) an end-to-end design concept, including cost scales, with self-consistent parameters, and 3) roadmaps and resource estimates for medium-term R\&D, including demonstrator facilities. The Design Study aims to be ready for the ``Targeted Panel'' envisioned by P5 later this decade \emph{(P5 Rec. 6)} that will make decisions on mid- and large-scale demonstrator facilities for future colliders. In order to prepare for the Targeted Panel and final design report, the Design Study is working towards unified metrics and benchmarks for the collider concept that will inform design choices. As an example, the \emph{geometric gradient} should be no less than 500 MeV/m in order to produce 5 TeV beams within the existing footprint of an LC facility. A complete set of metrics will be developed by the end of 2025 and an interim report that describes the collider design choices will be published by the end of 2026.

Lastly, we emphasize that a high-quality design study requires attention to detail, which is only possible through the effort of many individuals. Funding from the European and US High Energy Physics communities that supports postdocs and graduate students will facilitate the careful and specific work that is required by the 10 TeV Wakefield Collider Design Study. A modest investment today will result in increased clarity and better options for the HEP community when it comes time to decide on the pathway to a 10 TeV pCM collider.






\printbibliography

\newpage
\section*{Endorsers}
The live version of the list can be found at the 10 TeV Wakefield Collider Design Study input to ESPPU \href{https://indico.slac.stanford.edu/event/9691/overview}{Indico page.} The list below reflects a snapshot taken with 201 endorsers at 12:00 PDT on March 31, 2025.
\newline
\newline
Spencer Gessner,
Jens Osterhoff,
Carl A. Lindstr{\o}m,
Kevin Cassou,
Simone Pagan Griso,
Jenny List,
Erik Adli,
Brian Foster,
John Palastro,
Elena Donegani,
Moses Chung,
Mikhail Polyanskiy,
Lindsey Gray,
Igor Pogorelsky,
Gongxiaohui Chen,
Gianluca Sarri,
Brian Beaudoin,
Ferdinand Willeke,
David Bruhwiler,
Joseph Grames,
Yuan Shi,
Robert Szafron,
Angira Rastogi,
Alexander Knetsch,
Xueying Lu,
Douglas Storey,
Thomas Grismayer,
Michael Ehrlichman,
Maksim Kravchenko,
Christiane Scherb,
Vasyl Maslov,
Claudio Emma,
Daniel Kalvik,
Ivan Rajkovic,
Jean-Luc Vay,
Michael Peskin,
Tong Zhou,
Jeroen van Tilborg,
Patrick Meade,
Mario Galletti,
Carlo Benedetti,
Stewart Boogert,
Timothy Barklow,
Antonino Di Piazza,
Chaojie Zhang,
Yiheng Ye,
Stepan Bulanov,
Pratik Manwani,
Eric Esarey,
Cameron Geddes,
Livio Verra,
Laura Corner,
Oznur Apsimon,
Vera Cilento,
Allen Caldwell,
Rachel Margraf-O'Neal,
Qianqian Su,
Michael Litos,
Gerard Andonian,
Mark Hogan,
Carl Schroeder,
Kei Nakamura,
Jonathan Wood,
Anthony Vazquez,
Keegan Downham,
Davide Terzani,
Alexander Ody,
Sarah Schroeder,
Arianna Formenti,
Ulascan Sarica,
Thamine Dalichaouch,
Patric Muggli,
Rob Shalloo,
Francesco Massimo,
Alessandro Cianchi,
Jorge Vieira,
Felipe Pe\~na,
Chunguang Jing,
Francesco Filippi,
Pierre Drobniak,
Dillon Merenich,
Mariastefania De Vido,
Massimo Ferrario,
Robert Ariniello,
Andrei Seryi,
Weishuang Linda Xu,
Gemma Costa,
Jian Bin Ben Chen,
Lewis Kennedy,
Brendan O'Shea,
Kevin Langhoff,
Ole Gunnar Finnerud,
Nathan Majernik,
Lewis Boulton,
Dimitrios Ntounis,
Ankur Dhar,
Thomas Sch\"{o}rner,
Angelo Biagioni,
Ariel Schwartzman,
Kyoungchul Kong,
Roman P\"{o}schl,
Sophia Morton,
Gwanghui Ha,
Samuel Homiller,
John Farmer,
Csaba Balazs,
Jie Gao,
River Robles,
Emilio Nanni,
Graham Wilson,
Rogelio Tomas Garcia,
Bluemlein Johannes,
Juergen Reuter,
Angeles Faus Golfe,
Dirk Zerwas,
Jan Kalinowski,
Jonas Bj\"{o}rklund Svensson,
Ivanka Bozovic,
Mohammad Mahdi Altakach,
Marco Garten,
Johannes Braathen,
Francesco Giovanni Celiberto,
Carolina Amoedo,
Richard D'Arcy,
Edda Gschwendtner,
Joshua Gregory,
Pavel Karataev,
Patrick Koppenburg,
Leonhard Reichenbach,
Jan Pucek,
Matthew Wing,
Eduardo Granados,
Nigel Watson,
Tom Tong,
Ivo Schulthess,
Maria Enrica Biagini,
Hossein Saberi,
Jan Klamka,
Guoxing Xia,
Lars Reichwein,
Kyrre Ness Sjobak,
Andrea Renato Rossi,
Steffen Doebert,
Alban Sublet,
Samuel Norman,
Navid Vafaei-Najafabadi,
Ans Pardons,
Jonathan Bagger,
Paul Grannis,
Dongxing He,
Keisho Hidaka,
Alexander Pukhov,
Eugene Bulyak,
Leily Kiani,
Naveen Pathak,
William Li,
Pouya Asadi,
Anthony Gonsalves,
Eir Eline H{\o}rlyk,
Fanting Kong,
Lance Labun,
Mahek Logantha,
Ken Marsh,
Michele Bergamaschi,
Marcel Demarteau,
Mikael Berggren,
Maximilian Swiatlowski,
Eleni Vryonidou,
Oksana Chubenko,
Philippe Piot,
Katherine Fraser,
Siddharth Karkare,
Brigitte Cros,
Armando Valter Felicio Zuffi,
Nicole Hartman,
Aishik Ghosh,
Nelson Lopes,
Jaron Shrock,
Caterina Vernieri,
Tor Raubenheimer,
Marlene Turner,
Katharine Leney,
David Cooke,
Louis Forrester,
Thandikire Madula,
Anna Giribono,
Ou Labun,
Sebastien Corde,
Remi Lehe,
Noe Gonzalez,
Max Varverakis,
Matthias Fuchs,
Maxence Thevenet,
Sharon Perez,
Simon Knapen,
Aodhan McIlvenny,
Valentina Lee,
Claire Hansel,
Severin Diederichs,
Alexander Scheinker,
Rafi Hessami

\end{document}